\documentclass[conference]{IEEEtran}

\usepackage{amsmath,amssymb,amsthm}
\usepackage{enumerate}
\usepackage{stfloats}
\usepackage{comment}
\usepackage{subcaption}
\usepackage{siunitx}
\usepackage{mathtools}
\usepackage{accents,latexsym,cancel}
\usepackage{cite}

\usepackage[dvipsnames]{xcolor}
\definecolor{myPink}{RGB}{255,105,183}

\usepackage[T1]{fontenc}
\usepackage{graphics} 
\usepackage{epsfig} 
\usepackage[mathscr]{euscript}
\usepackage{algorithm}
\usepackage[noend]{algpseudocode}
\usepackage{bbm}
\makeatletter
\def\BState{\State\hskip-\ALG@thistlm}
\makeatother

\usepackage{tikz}
\usetikzlibrary{arrows,shapes,chains,matrix,positioning,scopes,patterns,calc,
decorations.markings,
decorations.pathmorphing,
}

\usepackage{pgfplots}
\pgfplotsset{compat=1.3}
\usepgflibrary{shapes}

\renewcommand{\epsilon}{\varepsilon}

\newcommand{\RNum}[1]{\uppercase\expandafter{\romannumeral #1\relax}}

\newcommand{\av}{\ensuremath{\mathbf{a}}}

\newcommand{\hv}{\ensuremath{\mathbf{h}}}

\newcommand{\pv}{\ensuremath{\mathbf{p}}}

\newcommand{\vv}{\ensuremath{\mathbf{v}}}

\newcommand{\wv}{\ensuremath{\mathbf{w}}}
\newcommand{\xv}{\ensuremath{\mathbf{x}}}
\newcommand{\yv}{\ensuremath{\mathbf{y}}}
\newcommand{\zv}{\ensuremath{\mathbf{z}}}

\newcommand{\gammav}{\ensuremath{\boldsymbol{\gamma}}}
\newcommand{\zerov}{\ensuremath{\boldsymbol{0}}}

\newcommand{\mcKa}{\ensuremath{\mathcal{K}_{\mathrm{a}}}}

\newcommand{\Ka}{\ensuremath{K_{\mathrm{a}}}}

\DeclareMathAlphabet{\mcl}{OMS}{cmsy}{m}{n}

\DeclareMathOperator*{\argmin}{arg\,min}
\DeclareMathOperator*{\argmax}{arg\,max}

\newlength\tikzwidth
\newlength\tikzheight

\definecolor{mycolor1}{rgb}{0.63529,0.07843,0.18431}%
\definecolor{mycolor2}{rgb}{0.00000,0.44706,0.74118}%
\definecolor{mycolor3}{rgb}{0.00000,0.49804,0.00000}%
\definecolor{mycolor4}{rgb}{0.87059,0.49020,0.00000}%
\definecolor{mycolor5}{rgb}{0.00000,0.44700,0.74100}%
\definecolor{mycolor6}{rgb}{0.74902,0.00000,0.74902}%

\newif\ifproof
\prooftrue

\def\fig_path{./Figures}
\IEEEoverridecommandlockouts
\begin{document}
\title{
Coded Compressed Sensing with Successive Cancellation List Decoding for Unsourced Random Access with Massive MIMO
}

%
\author{ Vamsi K. Amalladinne, Jean-Francois Chamberland, Krishna R. Narayanan\\
 Department of Electrical and Computer Engineering, Texas A\&M University
\thanks{
This material is based upon work supported, in part, by the National Science Foundation (NSF) under Grant CCF-1619085 and by Qualcomm Technologies, Inc., through their University Relations Program.}
}

\maketitle

\begin{abstract}
Unsourced random access (URA) is an increasingly popular communication paradigm attuned to machine driven data transfers in \textit{Internet-of-Things} (IoT) networks.
In a typical URA setting, a small subset of active devices within a very large population wish to transmit short messages to a central base station.
Originally defined for base stations equipped with a single antenna, the URA paradigm has recently been extended to practical scenarios involving base stations with a massive number of antennas by Fengler et al.
The proposed concatenated coding scheme therein utilizes a non-Bayesian sparse recovery algorithm coupled with the tree code introduced by Amalladinne et al.\ in the context of coded compressed sensing.
Currently, the existing MIMO implementation uses these two components in isolation.
This article introduces an enhanced successive cancellation list decoding style scheme that facilitates dynamic interactions between the sparse recovery algorithm and the tree decoder.
This modification can reduce the search space of the AD algorithm drastically; and it results in significant improvements both in terms of error performance and computational complexity.
Simulation results reveal that, for a system with 100 active users, the proposed decoder reduces the number of required antennas at the base station by 23\% to achieve a performance akin to the scheme by Fengler et al.
\end{abstract}


\section{Introduction}
\label{section:Introduction}

Conventional wireless systems have been designed to serve, primarily, traffic generated by humans, which is marked by sustained connections that persist over long periods of time. 
The emergence of communication between machines, \textit{Internet-of-Things} (IoT), and 5G is challenging existing infrastructures because wireless systems are ill-equipped to deal with short data bursts generated by a myriad of unattended devices. 
In contrast with human-centric data connections, machine-type communication (MTC) tends to be sporadic, with minute data payloads.
This shift may look benign.
However, the anticipated rise in device density, orders of magnitude beyond typical human population densities, makes it impractical to maintain a local catalogue of active devices, their buffer states, and their respective channel qualities.
This new reality, together with stringent requirements on delay, energy consumption, and spectral efficiency, demands a fundamental rethinking of wireless access, at least as it pertains to handling this rapidly growing traffic class.

This concerning situation has been broadly recognized by researchers who are currently exploring ways to ready wireless infrastructures for this evolving digital landscape \cite{bockelmann2016massive,mahmood2020white}.
Several technologies, including massive MIMO antenna systems and millimeter-wave radios have received much attention recently \cite{van2017massive, xiao2017millimeter}.
The renewed interest in random access schemes and grant-free communication are also motivated, partly, by an increasingly heterogeneous wireless traffic.

Unsourced random access (URA) is a novel multiple access paradigm introduced by Polyanskiy in~\cite{polyanskiy2017perspective} for uplink data transfers in dense wireless networks with sporadic activities.
From a modeling perspective, URA has emerged as a valuable communication paradigm for IoT applications and, as a consequence, it has become an active area of research~\cite{ordentlich2017low, vem2019user, amalladinne2020coded, pradhan2019sparseidma, pradhan2020polar, amalladinne2020enhanced, Giuseppe, amalladinne2020unsourced, calderbank2018chirrup, shyianov2020massive}.
The prime pursuit of these initiatives is the design of practical coding schemes for URA that admit low computational cost and perform close to the achievability bound in~\cite{polyanskiy2017perspective}, which is derived in the absence of complexity constraints.
Along these lines, we proposed coded compressed sensing (CCS) in~\cite{amalladinne2020coded}, an algorithmic framework that treats URA as a high dimensional support recovery problem and utilizes a divide-and-conquer approach for complexity reduction.
The key idea behind this scheme is to partition the payload corresponding to an active user into several sub-blocks and recover these components at the receiver using existing compressed sensing solvers.
These sub-blocks are enhanced with parity bits prior to the transmission phase, and a tree decoder that leverages this redundancy at the destination is employed for message disambiguation.
The CCS framework has attracted significant research attention, owing to its low implementation cost and good error performance.
Several extensions and enhancements have been proposed to advance the state-of-the-art in CCS~\cite{calderbank2018chirrup, Giuseppe, amalladinne2020enhanced, amalladinne2020unsourced, fengler2019non, shyianov2020massive}.
While some of these contributions aim to improve the performance of CCS in the standard URA setting \cite{calderbank2018chirrup, Giuseppe, amalladinne2020enhanced, amalladinne2020unsourced}, others extend this framework to more practical scenarios involving base stations equipped with a massive number of antennas~\cite{fengler2019non, shyianov2020massive}.

This article is very much aligned with this evolution as it seeks to combine two of these topics in advancing the state-of-the-art in wireless communications.
Specifically, we study the gains associated with multiple-antenna receivers in the context of URA.
In this sense, our work builds on a landmark contribution by Fengler et al.~\cite{fengler2019non} that showcases significant gains in activity detection when a massive number of antennas are present at the access point.
We demonstrate that, when combined with the enhanced successive cancellation list decoding (SCLD) style algorithm proposed in \cite{amalladinne2020enhanced}, the framework in \cite{fengler2019non} exhibits huge gains both in terms of error performance and computational complexity.

Throughout, we employ $\mathbb{Z}_+, \mathbb{Z}, \mathbb{C}, \mathbb{R}_+, \mathbb{R}$ to denote the set of all non-negative integers, integers, complex numbers,  non-negative real numbers, real numbers, respectively.
For any $n \in \mathbb{Z}_+$, we use $[n]$ to denote $\{0, 1, \ldots, n-1\}$.
Given binary vector $\xv$, $[\xv]_2$ refers to the integer whose radix-2 representation is $\xv$.

\section{System Model}
\label{section:SystemModel}

We consider a network with a total of $K_{\mathrm{tot}}$ users, among which $\Ka$ are active ($\Ka \ll K_{\mathrm{tot}}$).
We denote the set of active users by $\mcKa$.
User transmissions are synchronized through a beacon sent by the base station at the beginning of every frame and, as such, active devices are aware of frame boundaries.
The base station is equipped with $M \gg 1$ receive antennas, which are well separated to ensure that the spatial correlation among channels across antennas is negligible.
We adopt a block fading model akin to \cite{fengler2019massive}, where the MIMO channel is fixed for a coherence block of length $n$, and it is independent across coherence blocks.
Transmissions take place over $L$ such coherence blocks and, hence, the total number of channel uses spanned by a frame is given by $N=nL$.
Active user $k \in \mcKa$ transmits a codework $\xv_k \in \mathbb{C}^N$, which naturally assumes a block structure with $L$ blocks and $n$ channel uses within each block.
In the remainder of this article, for any vector $\vv$ that features this block structure, we use the overloaded notation $\vv(t,\ell)$ to denote the $t$th symbol of the $\ell$th block in $\vv$, and $\vv(\ell)$ to refer to the $\ell$th block of $\vv$.
The signal received at the base station during time instant $t$ of coherence block $\ell$ is then given by
\begin{align}
\label{eqn:systemModel}
\yv(t,\ell) = \sum_{k \in \mcKa} \xv_k(t,\ell) \hv_k(\ell) + \zv(t,\ell)
\quad t \in [n], \ell \in [L] 
\end{align}
where $ \hv_k(\ell) \sim \mathcal{CN}(0,\mathbf{I}_M)$ denotes the $M$-dimensional vector of small-scale fading coefficients corresponding to user $k$ as seen by the $M$ antennas at the base station, and additive term $\zv(t,\ell)$ represents circularly-symmetric complex white Gaussian noise, with mean zero and variance $\frac{N_0}{2}$ per dimension.

We refer to the collection of $B$-bit payloads associated with the active users as $\mathcal{W} = \{\wv_k, k \in \mcKa\}$.
Active user $k$ encodes its payload $\wv_k$ into a signal $\xv_k = f(\wv_k)$ and subsequently sends it over the MAC channel.
All active users employ a same codebook and, as such, the encoding function $f(\cdot)$ does not depend on the identity of transmitting user.
This is a defining characteristic of unsourced access~\cite{polyanskiy2017perspective}.
The signal transmitted by a user should respect the expected energy constraint $\|\xv_k\|_2^2 \le NP~\forall~k \in \mcKa$.
Accordingly, the energy-per-bit of this system is defined as $\frac{E_b}{N_0} = \frac{NP}{BN_0}$.

The base station is tasked with producing an estimate $\hat{\mathcal{W}} = \{\hat{\wv}_k\}$ of payloads transmitted by the active users using measurement vector $\yv$, as expressed in \eqref{eqn:systemModel}.
It has knowledge of the encoding function $f(\cdot)$ and of the second order statistics of the MIMO channels.
The cardinality of $\hat{\mathcal{W}}$ cannot exceed $\Ka$ to prevent the base station from making extraneous guesses.
System performance is measured using the per-user probability of error (PUPE) defined as~\cite{polyanskiy2017perspective}
\begin{align*}
P_{\mathrm{e}} = \textstyle \frac{1}{\Ka} \sum_{k \in \mcKa} \mathrm{Pr}(\wv_k \notin \hat{\mathcal{W}}).
\end{align*}
Our objective is to design a coding scheme that admits low computational complexity and achieves $P_{\mathrm{e}} \le \epsilon$, where $\epsilon$ is a target error probability.
The detection algorithm has to be non-coherent because the channel state information is unavailable either at the transmitter or at the receiver.

\section{Proposed Scheme}

The proposed scheme features a concatenated coding framework reminiscent of coded compressed sensing \cite{amalladinne2020coded, fengler2019non}.
The payload corresponding to an active user is divided into several sub-blocks and enhanced with redundancy using an outer tree code.
The coded sub-blocks are mapped into a signal of length $n$ using an inner encoder and transmitted during one coherence block.
The original algorithm proposed in \cite{fengler2019non} works as follows.
The destination first determines the collection of sub-blocks transmitted during a coherence block using a non-Bayesian covariance-based activity detection algorithm \cite{fengler2019non}.
Once sub-blocks corresponding to all coherence blocks are recovered, the redundancy employed by the tree encoder is used to disambiguate messages corresponding to different active users.
It is pertinent to note that in the original implementation \cite{fengler2019non}, the inner and outer decoding components operate independently of each other and as such, there are no dynamic interactions between them.
In this article, we propose an enhanced SCLD algorithm that allows the inner and outer decoders to operate in tandem.
This reduces the search space of the inner activity detection algorithm significantly by preemptively pruning the codebook through the list of active paths determined by the outer tree decoder.
We describe the encoder and decoder in detail below.

\subsection{Encoder}
\begin{figure}[tbh]
\centerline{\begin{tikzpicture}[
  font=\small, >=stealth', line width=0.75pt,
  infobits0/.style={rectangle, minimum height=6mm, minimum width=20mm, draw=black, fill=gray!10},
  infobits/.style={rectangle, minimum height=6mm, minimum width=12mm, draw=black, fill=gray!10},
  paritybits/.style={rectangle, minimum height=6mm, minimum width=8mm, draw=black, fill=lightgray!50}
]

\node[infobits0] (vb0) at (1,0) {$\wv(0)$};
\node[infobits] (vb1) at (2.6,0) {$\wv(1)$};
\node[paritybits] (vp1) at (3.6,0) {$\pv(1)$};
\node[infobits] (vb2) at (4.6,0) {$\wv(2)$};
\node[paritybits] (vp2) at (5.6,0) {$\pv(2)$};
\node[infobits] (vb3) at (6.6,0) {$\wv(3)$};
\node[paritybits] (vp3) at (7.6,0) {$\pv(3)$};
\draw[|-|] (0,-0.5) to node[midway,below] {$w_0$} (2,-0.5);
\draw[|-|] (2,-0.5) to node[midway,below] {$w_1$} (3.2,-0.5);
\draw[-|] (3.2,-0.5) to node[midway,below] {$p_1$} (4,-0.5);
\draw[|-|] (4,-0.5) to node[midway,below] {$w_2$} (5.2,-0.5);
\draw[-|] (5.2,-0.5) to node[midway,below] {$p_2$} (6,-0.5);
\draw[|-|] (6,-0.5) to node[midway,below] {$w_3$} (7.2,-0.5);
\draw[-|] (7.2,-0.5) to node[midway,below] {$p_3$} (8,-0.5);
\end{tikzpicture}}
\caption{This diagram illustrates the structure of CCS sub-blocks, with their information and parity bits.
Every sub-block is encoded separately before transmission within a slot.}
\label{figure:subblock}
\end{figure}
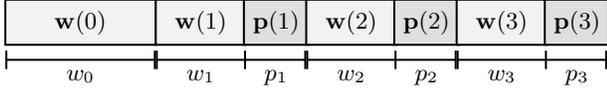

The $B$-bit payload $\wv$ corresponding to each active user is divided into $L$ sub-blocks with the $\ell$th sub-block containing $w_\ell$ information bits.
A total of $p_\ell$ parity-check bits are appended to sub-block $\ell$.
The first sub-block containing no parity-check bits, i.e., $p_0 = 0$.
This results in an encoded message reminiscent of Fig.~\ref{figure:subblock} with the $\ell$th sub-block consisting of $v_\ell = w_\ell+p_\ell$ bits.
The parity check bits $\pv(\ell)$ in sub-block $\ell$ are created by taking (random) linear combinations of all the information bits preceding sub-block $\ell$.
Mathematically, $\pv(\ell) = \sum_{j=0}^{\ell-1}\wv(j)\mathbf{G}_{j,\ell}$, where $\mathbf{G}_{j,\ell} \in \{0,1\}^{w_j \times p_\ell}$ is a binary Rademacher matrix and the computations are performed over the binary field to ensure parity-check bits are binary.
Overall, encoded messages take the form $\vv = \vv(0)~\vv(1)\cdots\vv(L-1)$, where the $\ell$th sub-block $\vv(\ell) = \wv(\ell) \pv(\ell)$.
The coded sub-blocks $\vv(\ell), \ell \in [L]$ are transmitted sequentially over various coherence blocks; they do not interfere with signals corresponding to other coherence blocks.

We turn to the description of the inner encoder, which maps coded sub-blocks to signals transmitted over the MAC channel.
Let $\mathbf{A}(\ell) \in \mathbb{C}^{n \times 2^{v_\ell}} = [\av_0(\ell),\cdots,\av_{2^{v_\ell}-1}(\ell)]$ denote the common codebook used by active users to transmit their coded sub-blocks during coherence block $\ell$. 
The columns of $\mathbf{A}(\ell)$ are normalized to ensure $\|\av_j(\ell)\|_2^2 = nP$ for $j \in [2^{v_\ell}]$.
The coded sub-block $\vv(\ell)$ corresponding to an active user is encoded into a column of $\mathbf{A}({\ell})$ using the bijective function $\vv(\ell) \mapsto [\vv(\ell)]_2$.
More specifically, the sent signal is obtained by mapping binary sequenced $\vv(\ell)$ to the $[\vv(\ell)]_2$th column of $\mathbf{A}(\ell)$.
Active users then transmit the chosen columns of $\mathbf{A}(\ell)$ as codewords during coherence slot $\ell$ over the MAC channel.

\subsection{Decoder}
The decoder features two components: an inner decoder that utilises a covariance-based activity detection algorithm to determine the transmitted sub-blocks; and an outer tree decoder that disambiguates messages across sub-blocks.
In the remainder of this section, we first review the framework introduced in \cite{fengler2019non}, where the inner and outer decoders operate sequentially, in isolation from each other.
Then, we describe the algorithmic enhancement based on successive cancellation list decoding that allows dynamic interactions between the inner activity detection algorithm and the outer tree decoder.

\begin{figure}[tbh]
\centerline{\begin{tikzpicture}[
  font=\small, >=stealth', line width = 0.75pt,
  wblock/.style={rectangle, minimum height=2mm, draw=black, fill=gray!10},
  pblock/.style={rectangle, minimum height=2mm, draw=black, fill=gray!40}
]

\foreach \y in {0, 0.5, 1, 1.5} {
    \foreach \x in {0, 0.5, 1, 1.5, 2, 2.5, 3, 3.5, 4, 4.5} {
        \draw (\x,\y) to (\x,\y+0.2) to (\x-0.1,\y+0.3) to (\x+0.1,\y+0.3) to (\x,\y+0.2);
     }
}

\draw[|-|] (-0.15, 2.15) -- node[above] {$M$ antennas} (4.65, 2.15);
\draw[->] (-0.35, 1.8) -- node[above,rotate=90] {Time} (-0.35,0);

\end{tikzpicture}}
\caption{Matrix $\mathbf{Y}(\ell)$ corresponding to coherence block~$\ell$ has $n$ rows, one for every time instant~$t$, and $M$ columns, one for every antenna.}
\label{figure:MIMO}
\end{figure}
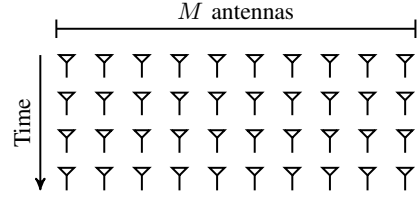

Suppose $\{i_k(\ell):k \in \mcKa\}$ denotes the indices corresponding to the columns picked by the active users for coherence block $\ell$.
Then, the signal received at the base station during coherence block $\ell$ can be written in matrix equation form as
\begin{equation} \label{eqn:subblock}
\begin{split}
\mathbf{Y}(\ell) &= \sum_{k \in \mcKa} \av_{i_k(\ell)}(\ell) {\hv_k(\ell)}^\top + \mathbf{Z}(\ell) \\
&= \mathbf{A}(\ell)\mathbf{\Gamma}(\ell)\mathbf{H}(\ell) + \mathbf{Z}(\ell).
\end{split}
\end{equation}
In the above equation, $\mathbf{H}(\ell) \in \mathbb{C}^{2^{v_\ell} \times M}$ has entries drawn i.i.d.\ from $\mathcal{CN}(0,1)$,
and $\mathbf{\Gamma}(\ell)$ is a diagonal matrix that indicates the indices of the columns transmitted by active users during coherence block~$\ell$. 
This matrix can be succinctly expressed as $\mathbf{\Gamma}(\ell) = \mathrm{diag}(\gammav(\ell)) = \mathrm{diag}(\gamma_0(\ell), \ldots, \gamma_{2^{v_\ell}-1}(\ell))$, where
\begin{align*}
 \gamma_i(\ell) = \begin{cases} 1~&\text{if}~i \in \{i_k(\ell): k \in \mcKa\} \\
 0~&\text{otherwise.}
 \end{cases}
\end{align*}
Additive noise $\mathbf{Z}(\ell) \in \mathbb{C}^{n \times M}$ consists of i.i.d. Gaussian noise samples with mean 0 and variance $\frac{N_0}{2}$ per dimension.
It may be useful to emphasize that the $t$th row ($t \in [n]$) of $\mathbf{Y}(\ell)$ captures the signal received at the $M$ antennas of base station during time instant $t$ of coherence block $\ell$.
Figure~\ref{figure:MIMO} offers an illustration of the structure of $\mathbf{Y}(\ell)$.

The columns of $\mathbf{Y}(\ell)$ are independent and identically distributed.
Column~$i$, where $i \in [M]$, has complex Gaussian distribution, with $\mathbf{y}_i(\ell) \sim \mathcal{CN}(\zerov,\mathbf{\Sigma}_\ell)$.
We can therefore express covariance matrix $\mathbf{\Sigma}_\ell$ as
\begin{equation} \label{equation:CovarianceMatrix}
\begin{split}
&\mathbf{\Sigma}_\ell
= \mathrm{E} \left[ \mathbf{y}_i(\ell) \mathbf{y}_i(\ell)^H \right]
= \frac{1}{M} \mathrm{E} \left[ \mathbf{Y}(\ell) \mathbf{Y}(\ell)^H \right]
\end{split}
\end{equation}
which reduces to $\mathbf{\Sigma}_\ell = \mathbf{A}(\ell)\mathbf{\Gamma}(\ell)\mathbf{A}(\ell)^H + N_0\mathbf{I}_n$ after simplification.
We note that the simpler expression leverages the facts: $\mathbf{\Gamma}(\ell) \mathbf{\Gamma}(\ell)^H = \mathbf{\Gamma}(\ell)$, and the columns of $\mathbf{H}(\ell)$ are uncorrelated.
The latter property allows us to write
\begin{equation*}
\mathrm{E} \left[ \mathbf{H}(\ell) \mathbf{H}(\ell)^H \right]
= \sum_{m \in [M]} \mathrm{E} \left[ \mathbf{H}(\ell, m) \mathbf{H}(\ell, m)^H \right]
= M \mathbf{I}_n
\end{equation*}
because the cross terms vanish.
This, along with the fact that the additive noise is a sequence of i.i.d.\ components, produced the structure in \eqref{equation:CovarianceMatrix}.
The covariance of $\mathbf{Y}(\ell)$ can be estimated using empirical averaging, with $\hat{\mathbf{\Sigma}}_{\mathbf{Y}(\ell)} = \frac{1}{M}\mathbf{Y}(\ell)\mathbf{Y}(\ell)^H$.
This estimator is consistent and, when the number of antennas is large, it performs well.

Under our problem formulation, identifying the sub-blocks sent during block~$\ell$ then becomes equivalent to detecting the locations of non-zero entries in the principal diagonal of $\mathbf{\Gamma}(\ell)$.
We employ the activity detection algorithm presented in \cite{fengler2019non} to accomplish this task.
Specifically, the activity detection algorithm seeks to obtain a constrained maximum-likelihood (ML) estimate of $\gammav(\ell)$ given by
\begin{align}
\label{optimizationProblem}
\gammav^*(\ell) = \argmax_{\gammav(\ell) \in \mathbb{R}_+^{2^{v_\ell}}} \mathcal{L} (\gammav(\ell))
\end{align}
where $\mathcal{L} (\gammav(\ell))$ denotes the log-likelihood function of the conditional distribution $p(\mathbf{Y}(\ell) | \gammav(\ell))$.
As a side note, we stress, in the original problem we consider, the ML search should be over discrete, sparse vectors whose entries are confined to $\{ 0, 1 \}$.
Yet, the relaxed approach in \eqref{optimizationProblem}, with its search over the non-negative orthant, is one of many tactics adopted in~\cite{fengler2019non} to keep the solution tractable.
We can compute the $\argmax$ of $\log p(\mathbf{Y}(\ell) | \gammav(\ell))$ as follows
\begin{equation}
\begin{split}
\argmax &\; \mathcal{L} (\gammav(\ell)) = \argmax \log p(\mathbf{Y}(\ell) | \gammav(\ell)) \\ 
&= \argmax \frac{1}{M} \sum_{i=1}^{M} \log p(\mathbf{y}_i(\ell) | \gammav(\ell)) \\
&= \argmin \left( \log |\mathbf{\Sigma}_\ell| + \operatorname{trace} \left( \mathbf{\Sigma}_\ell^{-1} \hat{\mathbf{\Sigma}}_{\mathbf{Y}(\ell)} \right)\right).
\end{split}
\end{equation}
The optimization problem \eqref{optimizationProblem} is non-convex and, in general, difficult to solve.

We adopt the iterative coordinate descent algorithm found in \cite{fengler2019non} to get an approximate solution to \eqref{optimizationProblem}.
For the sake of completeness, we reproduce the algorithmic details below, in Algorithm~\ref{alg:activity}.
Therein, we use $\mathcal{S}_\ell$ to denote the list of diagonal indices to perform descent over; and, for the time being, this set is simply $\mathcal{S}_\ell = [2^{v_\ell}]$.
We delay the treatment of the implementation twist necessary to facilitate dynamic interactions between the activity detection algorithm and the outer tree decoder until after we establish this foundation.
Ultimately, the estimated list of coded sub-blocks transmitted during coherence block $\ell$ is obtained by picking the indices corresponding to the $\Ka+\delta$ largest entries in $\gammav^*(\ell)$, where $\delta \ge 0$ is a small integer.

\begin{algorithm}[t!]
\caption{Activity Detection via Coordinate Descent}\label{alg:activity}
\begin{algorithmic}[1]
\State \textbf{Inputs}: Sample covariance $\hat{\mathbf{\Sigma}}_{\mathbf{Y}(\ell)} = \frac{1}{M}\mathbf{Y}(\ell)\mathbf{Y}(\ell)^H$
\State \textbf{Initialize}: $\mathbf{\Sigma}_\ell = N_0 \mathbf{I}_n, \gammav(\ell) = 0$
\For {$i=1,2,\ldots$} 
	\For {$k \in \mathcal{S}_\ell$}
		\State Set $d^* = \frac{\av_k(\ell)^H \mathbf{\Sigma}_\ell^{-1} (\hat{\mathbf{\Sigma}}_{\mathbf{Y}(\ell)}\mathbf{\Sigma}_\ell^{-1} - \mathbf{I}_n)\av_k(\ell)} {(\av_k(\ell)^H \mathbf{\Sigma}_\ell^{-1} \av_k(\ell))^2}$
		\State Update $\gamma_k(\ell) \gets \max \{ \gamma_k(\ell) + d^*, 0 \}$
		\State Update $\mathbf{\Sigma}_\ell^{-1} \gets \mathbf{\Sigma}_\ell^{-1} - \frac{d^*\mathbf{\Sigma}_\ell^{-1}\av_k(\ell)\av_k(\ell)^H\mathbf{\Sigma}_\ell^{-1}}{1 + d^*\av_k(\ell)^H\mathbf{\Sigma}_\ell^{-1}\av_k(\ell)}$
	\EndFor
\EndFor
\State \textbf{Output}: Estimate $\gammav(\ell)$
\end{algorithmic}
\end{algorithm}

Once the transmitted sub-blocks are determined by the inner decoder, the outer tree decoder originally proposed in \cite{amalladinne2020coded} works across sub-blocks to stitch information corresponding to an active user together.
We briefly describe the tree decoder below and refer the reader to \cite{amalladinne2020coded} for a comprehensive description and details regarding performance analysis.
The destination constructs a decoding tree for every potentially transmitted message sequence.
The root node for a decoding tree corresponds to a coded sub-block returned by the activity detection algorithm for coherence slot $0$.
The decoder then computes the parity check bits $\pv(1)$ resulting from the considered root node.
All the sub-blocks at level $1$ containing parity check bits that match $\pv(1)$ are considered active and attached to the root node and the remaining ones are discarded.
Subsequently, parity check bits $\pv(2)$ are computed for every active path emerging from the root node and sub-blocks at slot $2$ that match these parity check bits are retained in the decoding tree while others are discarded.
This process continues until slot $L-1$ is reached.
If only one path survives all the stages, it is deemed a valid message.
In all the other cases, a decoding failure is declared by the tree decoder; the corresponding root node is declared invalid and, subsequently, discarded.

\textbf{Successive Cancellation List Decoding:}
The proposed SCLD scheme parallels an idea introduced in \cite{amalladinne2020enhanced}, but adapted to the problem at hand, which involves multiple receive antennas at the base station.
The key realization behind this successive cancellation scheme is that, upon performing fragment recovery using the AD algorithm at slot $\ell-1$, the outer decoder can identify the list of all active paths until stage $\ell-1$ by running several instances of the tree decoding algorithm described above, one instance for every root fragment.
Furthermore, the collection of permissible parity patters $\mathcal{P}_{\ell}$ that emanate from all the active paths until stage $\ell-1$ can be pre-computed and supplied as side-information to the AD algorithm at slot $\ell$.
Parity patterns that are not present in $\mathcal{P}_{\ell}$ cannot be reached by any active path and would eventually be discarded by the tree decoder.
Hence, the columns of $\mathbf{A}(\ell)$ corresponding to these inadmissible parity patters can be pruned (cancelled) before an attempt is made to recover active sub-blocks at slot $\ell$ by the inner AD algorithm.
This simplifies the fragment recovery process at slot $\ell$ significantly by reducing the search space of the AD algorithm to a subset $\mathcal{S}_\ell \subseteq [2^{v_\ell}]$ of columns in $\mathbf{A}(\ell)$.
Mathematically, the collection of effective columns in $\mathbf{A}(\ell)$ given the past observations $\mathcal{P}_\ell$ can be expressed as
\begin{equation*}
\mathcal{S}_\ell = \left\{ [\wv(\ell)\pv(\ell)]_2 \middle| \wv(\ell) \in \{0,1\}^{{w_\ell}}, \pv(\ell) \in \mathcal{P}_\ell \right\} .
\end{equation*}
It is easy to see from the above formulation that the effective number of columns in $\mathbf{A}(\ell)$ becomes $|\mathcal{S}_\ell| = 2^{w_\ell}|\mathcal{P}_\ell|$ which can be much smaller than the total number of columns $2^{v_\ell}$.
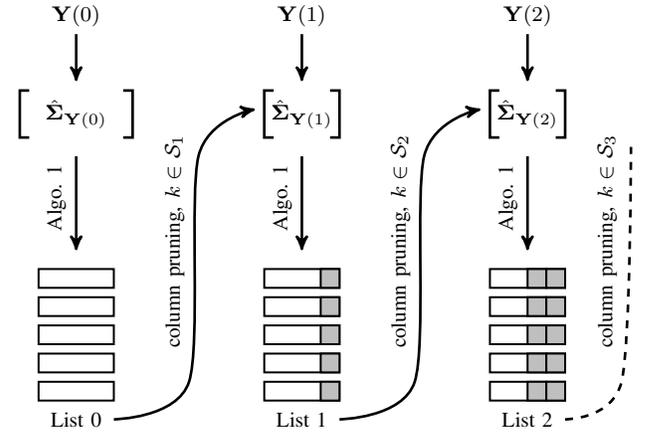
\begin{figure}[tbh]
\centerline{\begin{tikzpicture}
[font=\footnotesize, draw=black, line width=0.75pt,>=stealth',
sub0/.style={rectangle, draw, inner sep=0pt, minimum width=10mm, minimum height=2.5mm},
parity/.style={rectangle, draw, fill=lightgray, inner sep=0pt, minimum size=2.5mm}]

\node (cs1) at (0,7) {$\mathbf{Y}(0)$};
\node (cs2) at (3,7) {$\mathbf{Y}(1)$};
\node (cs3) at (6,7) {$\mathbf{Y}(2)$};

\foreach \v in {0,3,6} {
  \draw[->, line width=1pt]  (\v,5.125) -- node[above, rotate=90,]{\, Algo.~1} (\v,3.875);
  \draw[->, line width=1pt]  (\v,6.75) -- (\v,6.125);
}

\foreach \v in {0} {
  \draw[line width=1pt] (\v-0.625,6) -- (\v-0.75,6) -- (\v-0.75,5.375) -- (\v-0.625,5.375);
  \draw[line width=1pt] (\v+0.625,6) -- (\v+0.75,6) -- (\v+0.75,5.375) -- (\v+0.625,5.375);
}
\node at (0,5.6875) {$\hat{\mathbf{\Sigma}}_{\mathbf{Y}(0)}$};

\foreach \v in {3,6} {
  \draw[line width=1pt] (\v-0.375,6) -- (\v-0.5,6) -- (\v-0.5,5.375) -- (\v-0.375,5.375);
  \draw[line width=1pt] (\v+0.375,6) -- (\v+0.5,6) -- (\v+0.5,5.375) -- (\v+0.375,5.375);
}
\node at (3,5.6875) {$\hat{\mathbf{\Sigma}}_{\mathbf{Y}(1)}$};
\node at (6,5.6875) {$\hat{\mathbf{\Sigma}}_{\mathbf{Y}(2)}$};

\foreach \p/\c in {3.50/1, 3.125/2, 2.75/3, 2.375/4, 2/5} {
  \node[sub0] (subcs1\c) at (0,\p) {};
  \node[sub0] (subcs2\c) at (3,\p) {};
  \node[parity] (parity0\c) at (3.375,\p) {};
  \node[sub0] (subcs3\c) at (6.00,\p) {};
  \node[parity] (parity1\c) at (6.125,\p) {};
  \node[parity] (parity2\c) at (6.375,\p) {};
}

\node (list1) at (0,1.625) {List~0};
\node (list2) at (3,1.625) {List~1};
\node (list3) at (6,1.625) {List~2};

\draw [line width=1pt,->] plot[smooth, tension=.5] coordinates {(0.5,1.625) (1.5,2.125) (1.625,5.125) (2.375,5.75)};
\draw [line width=1pt,->] plot[smooth, tension=.5] coordinates {(3.5,1.625) (4.5,2.125) (4.625,5.125) (5.375,5.75)};
\draw [line width=1pt,dashed] plot[smooth, tension=.5] coordinates {(6.5,1.625) (7.25,2.125) (7.375,5.25)};

\node[rotate=90] (prune1) at (1.3125,4) {column pruning, $k \in \mathcal{S}_1$};
\node[rotate=90] (prune2) at (4.3125,4) {column pruning, $k \in \mathcal{S}_2$};
\node[rotate=90] (prune3) at (7.0625,4) {column pruning, $k \in \mathcal{S}_3$};
\end{tikzpicture}}
\caption{This diagram illustrates the reduction in the search space of the inner activity detection algorithm afforded by successive cancellation list decoding.}
\label{figure:eCCS}
\end{figure}

\section{Simulation Results and Discussion}
\label{section:SimulationResults}
We consider an uplink transmission scenario with $\Ka$ active users and a base station equipped with $M$ receive antennas, where $25 \le \Ka \le 150$ and $25 \le M \le 125$.
The transmission frame consists of $L = 32$ coherence blocks and each coherence block has $n = 100$ complex channel uses.
Therefore, the total number of complex channel uses in a frame is given by $N = nL = 3200$.
The \textit{energy-per-bit} of the system is set to $\frac{E_b}{N_0} = 0$ dB.
The payload size of each active user is $B = 96$ bits.
All the coded sub-blocks are of length $12$ bits, i.e., $w_\ell + p_\ell = 12~\forall~\ell \in [L]$.
We choose the parity profile $(p_0,p_1,\cdots,p_{L-1}) = (0,9,9,\cdots,9,12,12,12)$ throughout these simulations.
The columns of codebook matrix $\mathbf{A}(\ell)$ are chosen uniformly at random from a sphere of radius $\sqrt{nP}$.
These simulation parameters are aligned with \cite{fengler2019non} to ensure a fair comparison.

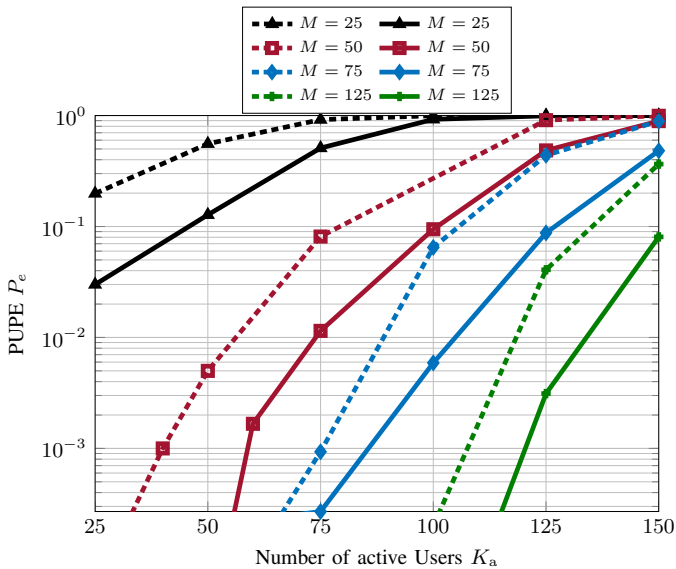
\begin{figure}[!ht]
\centering
\resizebox{0.5\textwidth}{!}{\begin{tikzpicture}

\begin{semilogyaxis}[%
font=\small,
width=10cm,
height=7.5cm,
xmin=25,
xmax=150,
xtick = {25,50,...,150},
xlabel={Number of active Users $\Ka$},
xmajorgrids,
ymin=0.00027,
ymax=1,
ytick = {0.0001,0.001,0.01,0.1,1},
ylabel={PUPE $P_e$},
ymajorgrids,
yminorgrids,
legend style={at={(0.5,1.275)},anchor=north,draw=black, fill=white, legend cell align=left,font=\scriptsize, legend columns=2}
]

\addplot [color=black,densely dashed,line width=2.0pt,mark=triangle,mark size=1.5pt,mark options={solid}]
  table[row sep=crcr]{
25 0.198400 \\
50 0.5566 \\
75 0.9188 \\
100 0.997 \\
125 0.9992 \\
150 1 \\
};
\addlegendentry{$M=25$};

\addplot [color=black,line width=2.0pt,mark=triangle,mark size=1.5pt,mark options={solid}]
  table[row sep=crcr]{25 0.03 \\
50 0.1274 \\
75 0.5096 \\
100 0.9247 \\
125 0.99416 \\
150 0.999467 \\
};
\addlegendentry{$M=25$};

\addplot [color=mycolor1,densely dashed,line width=2.0pt,mark=square,mark options={solid}]
  table[row sep=crcr]{
25 0\\
32 0.000208\\
40 0.001000\\
50 0.005 \\
75 0.081200 \\
125 0.909120\\
150 0.995333 \\
};
\addlegendentry{$M=50$};

\addplot [color=mycolor1, line width=2.0pt,mark=square,mark options={solid}]
  table[row sep=crcr]{
25 0\\
50 0\\
55 0.000182\\
60 0.001667\\
75 0.011467\\
100 0.094500\\
125 0.485120\\
150 0.891133\\
};
\addlegendentry{$M=50$};

\addplot [color=mycolor2,densely dashed,line width=2.0pt,mark=diamond,mark options={solid}]
  table[row sep=crcr]{
  25 0\\
  50 0\\
  65 0.00021\\
  75 0.000933\\
  100 0.064800\\
  125 0.438560 \\
  150 0.889600\\
};
\addlegendentry{$M=75$};

\addplot [color=mycolor2,line width=2.0pt,mark=diamond,mark options={solid}]
  table[row sep=crcr]{25 0 \\
  50 0.0002 \\
  75 0.00027\\
  100 0.0059 \\
  125 0.0876 \\
  150 0.481667 \\
};
\addlegendentry{$M=75$};

\addplot [color=mycolor3,densely dashed,line width=2.0pt,mark=+,mark options={solid}]
  table[row sep=crcr]{
  100 0.0002 \\
  125 0.040720 \\
  150 0.365333\\
};
\addlegendentry{$M=125$};

\addplot [color=mycolor3,solid,line width=2.0pt,mark=+,mark options={solid}]
  table[row sep=crcr]{100 0 \\
  115 0.000261 \\
  125 0.003120\\
  150 0.080667 \\
};
\addlegendentry{$M=125$};

\end{semilogyaxis}

\end{tikzpicture}%
}
\caption{This graph showcases performance comparison between the original algorithm in \cite{fengler2019non} and the enhanced SCLD scheme proposed in this article. Dashes lines represent the performance of the original algorithm in \cite{fengler2019non} and solid lines represent the performance of the enhanced SCLD scheme proposed in this article.}
\label{fig:sim_results1}
\end{figure}

In Fig.~\ref{fig:sim_results1}, the per-user probability  of error $P_e$ is plotted for different values of $\Ka$ and $M$.
As expected, the proposed SCLD scheme substantially outperforms the original scheme in \cite{fengler2019non}.
Specifically, when the number of active users is $100$, the proposed SCLD enhancement reduces the number of required antennas at the base station by 23\% to achieve a performance akin to the scheme in \cite{fengler2019non}.
Fig.~\ref{fig:sim_results1} showcases a comparison of the average run-times between the two algorithms.
The ratio of average run-times of the enhanced SCLD algorithm and the original algorithm in \cite{fengler2019non} is plotted as a proxy for computational complexity for different values of $\Ka$.
The proposed scheme demonstrates significant complexity reduction and the gain is more pronounced in the regime where the number of active users is low.

\begin{figure}[!ht]
\centering
\resizebox{0.5\textwidth}{!}{\begin{tikzpicture}

\begin{axis}[%
font=\small,
width=10cm,
height=7.5cm,
xmin=25,
xmax=150,
xtick = {25,50,...,150},
xlabel={Number of active Users $\Ka$},
xmajorgrids,
ymin=0,
ymax=0.4,
ytick = {0,0.1,...,0.4},
ylabel={Ratio of average run-times},
ymajorgrids,
legend style={at={(1,1)},anchor=north east,draw=black,fill=white,legend cell align=left}
]

\addplot [color=red,solid,line width=2.0pt,mark=diamond,mark options={solid}]
  table[row sep=crcr]{
  25 0.065355\\
  50 0.1080\\
  75 0.1522\\
  100 0.2022\\
  125 0.2464\\
  150 0.2914\\
};
\addlegendentry{Ratio of average run-times};






\end{axis}

\end{tikzpicture}%
}
\caption{This graph showcases comparison between the average run-times of the original algorithm in \cite{fengler2019non} and the enhanced SCLD scheme proposed in this article when the number of receive antennas is $M=50$. The ratio of average run-times between the proposed scheme and the original scheme in \cite{fengler2019non} is plotted as a function of the number of active users $\Ka$.}
\label{fig:sim_results2}
\end{figure}
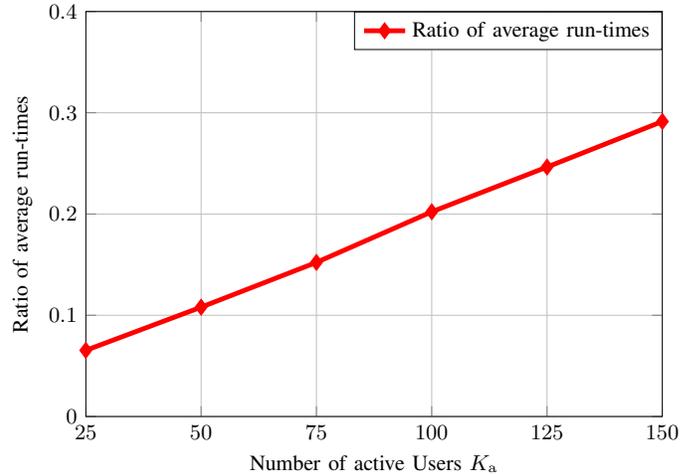

\section{Conclusion}
\label{sec:Conclusion}
This article advances state-of-the-art in unsourced access with a massive MIMO receiver.
The algorithmic framework considered in this article is reminiscent of the concatenated coding scheme employed in other related works in this area.
The proposed enhanced SCLD scheme leverages pertinent information provided by the outer decoder to reduce the search space of inner activity detection algorithm.
This leads to significant gains both in terms of error performance and computational complexity of decoding.
Furthermore, simulation results reveal that the number of antennas that need to be installed at the base station can be reduced significantly to achieve a performance level akin to the state-of-the-art scheme.



\bibliographystyle{IEEEbib}
\bibliography{IEEEabrv,MACcollison}


\end{document}